\documentclass[conference]{IEEEtran}
\pdfoutput=1
\usepackage{fixltx2e}
\usepackage{cite} \usepackage[pdftex]{graphicx}
\usepackage[cmex10]{amsmath}
\usepackage[caption=false,font=normalsize]{subfig}
\usepackage{url}
\usepackage{multirow}
\usepackage[utf8x]{inputenc} \usepackage[T1]{fontenc}
\usepackage{hyperref}
\usepackage{cleveref}
\usepackage{booktabs}
\graphicspath{{figs/}}
\newcommand{\etal}{et al.}
\newcommand{\lfm}{last.fm}
\newcommand{\Lfm}{Last.fm}
\newcommand{\xhdr}[1]{\vspace{2mm} \noindent{\bf #1. }}

\newcommand{\specialcell}[2][c]{%
  \begin{tabular}[#1]{@{}c@{}}#2\end{tabular}}

\begin{document}

\title{The Geographic Flow of Music}
\author{\IEEEauthorblockN{Conrad
    Lee and P\'{a}draig Cunningham}
    \IEEEauthorblockA{Clique Research Cluster\\
                      University College Dublin\\
                      8 Belfield Office Park, Clonskeagh\\
                      Dublin 4, Ireland\\
                      Tel: +353 1 716 5346 \\
                      Email: conradlee@gmail.com, padraig.cunningham@ucd.ie}}

\maketitle

\begin{abstract}
  The social media website \lfm{} provides a detailed snapshot of what
  its users in hundreds of cities listen to each week. After suitably
  normalizing this data, we use it to test three hypotheses related to
  the geographic flow of music. The first is that although many of 
  the most popular artists are listened to around the world, music
  preferences are closely related to nationality, language, and
  geographic location. We find support for this hypothesis, with a
  couple of minor, yet interesting, exceptions. Our second hypothesis
  is that some cities are consistently early adopters of new music
  (and early to snub stale music). To test this hypothesis, we adapt a
  method previously used to detect the leadership networks present
  in flocks of birds. We find empirical support for the claim that a
  similar leadership network exists among cities, and this finding is the main
  contribution of the paper. Finally, we test the hypothesis that large
  cities tend to be ahead of smaller cities--we find only
  weak support for this hypothesis.
\end{abstract}

\IEEEpeerreviewmaketitle

\section{Introduction}
The question of how information and preferences spread through social
networks has a long and rich history. The topic became an active field of study
just after World War Two \cite{lazarsfeld1948,coleman1957,moreno1953,festinger1950};
this early work produced, for example, the seminal \textit{two-step
  flow of communication} hypothesis, which states that ``ideas often
flow from radio and print to opinion leaders, and from these to the
less active sections of the population'' \cite{katz1957}.
In the 1970s, Mark Granovetter contributed prominent ideas to the
field, including the hypothesis that members of tightly-knit social
groups have largely duplicate information, and rely on
acquaintanceships with members of other groups to gain access to novel
information \cite{granovetter1973}.

More recently, detailed logs of digital communication
have enabled these hypotheses
to be tested on datasets that are much larger than was feasible only a
decade earlier. For example, in \cite{bakshy2012}, Bakshy \etal{}
subject 250 million facebook users to a controlled experiment in order
to measure the role that facebook friends play in influencing the diffusion of
information, finding that while a user's most active relationships are
individually the most influential, the overall effect of less active
relationships in spreading novel information is stronger. Additional examples of recent
significant work includes the worldwide spread of e-mail chain letters
\cite{libennowell2008}, the analysis of a massive worldwide instant
messaging dataset \cite{leskovec2008}, and the spread of information
through the blogosphere \cite{gruhl2004,leskovec2009}.


Here, we investigate hypotheses related to the geographic flow of
preferences in music. Our main contribution is to formalize and answer
the following question:
if one considers the month-by-month change in the aggregate musical
preferences of cities, are some cities consistently ahead of others?
In other words, can we find that some cities are leaders and others
are followers?

Our enquiry into the geographic distribution of musical preferences is
structured as follows.  We begin by describing the data, a world-wide log of
listening habits recorded by \lfm, as well as various
pre-processing and normalization steps in \cref{sec:data}.
Next, in \cref{sec:statistics}, we measure how regional musical preferences are,
finding that although many of the most
popular artists are popular all around the world, there are
nonetheless well-defined clusters of cities that are closely related
to nationality, language, and geographic distance.

In \cref{sec:vector-analysis}, we move on to our main contribution: an
analysis of the \textit{dynamics} of music
preferences. We adapt a methodology
previously used to find leadership in pigeon flocks \cite{Nagy2010} to
detect whether some cities consistently follow others. At a high
level, this methodology involves looking at every dyad (pair of nodes) and running
a test to see whether the time-lagged correlation is larger in one
direction than another.  We observe that when we put all of these
directed pairs together, the resulting networks are nearly
acyclic, a strong indicator that the geographic flow of music
has a clear direction, i.e., hierarchical structure \cite{mones2012}.

Recently there has been much excitement surrounding the observation
that productivity, efficiency, and innovation all scale super-linearly
with the size of a city
\cite{bettencourt2011,bettencourt2007,arbesman2011}. This line of
reasoning suggests the hypothesis that larger cities should also be
more up to date on the latest and greatest music. We wrap up our
inquiry into the spread of music in \cref{sec:superlinearity} by
testing the hypothesis that leadership is also correlated with the
size of a city.


\section{Data: preprocessing \& normalization}

\label{sec:data}
\Lfm{} is a service based around collecting data on the listening
habits of its users. Users install a plug-in on their audio players
such as iTunes or Winamp which keeps track of the songs that the user
listens to, either on his computer or external device (e.g., an
iPod). The plug-in uploads this information to the \lfm{} database,
giving the service a log of what its users listen to.  In 2011 alone,
\lfm{} received 11 billion such notifications (called ``scrobbles'' by
last.fm), and since the service began in 2003 it has received 61
billion.\footnote{According to \lfm's blog post at
  http://blog.last.fm/2012/01/16/building-best-of-2011} \Lfm{} uses
this information in various ways, for example, to compare the
similarity of two users' musical taste, to recommend music, and to
create a profile page.

\xhdr{Creating \textit{listen matrices}}\Lfm{} aggregates this data
into weekly charts for over 200 metropolitan areas around the world,
and makes the data behind these charts accessible through a public
API. For every week and each city, the \lfm{} API indicates the number
of unique listeners that each of that city's top 500 artists
had. Thus, for each week we have a matrix; in this matrix every city
is a row vector with 500 non-zero elements, and each column represents
an artist.  Because not all cities have the same top 500 artists, the
matrix has more than 500 columns and a large number of
zero-valued elements.  Thus, a non-zero entry in this matrix at
position $i,j$ is a positive integer indicating the number of unique
users from city $i$ who listened to artist $j$ that week. Zero-valued
entries indicate that the artist had either no listeners, or that it
was not among the 500 most popular artists in the city that week. At
the time of data collection in late 2011, these charts were available
for 153 weeks.

Because not all \lfm{} users are active every week, a single week's
chart can be thought of as a sample of listening preferences among
\lfm{} users.  In cities that have relatively few users, the variance
associated with this sample becomes large, indicating noise. We find that
we can reduce this noise by summing up the matrices associated with
four consecutive weeks together.  This effectively increases the
sample size for each entry in the city-artist matrix described above. For
this reason, in all of the analysis below, we aggregate our data
using a ``sliding window'' where the width of the window is four
weeks, and the window slides in one-week steps. We call the matrices
associated with these four-week periods \textit{listen matrices}.

\xhdr{Normalizing listen matrices} Consider the toy example presented
in \cref{fig:vec-analysis}(A). In this scenario, we imagine there are
only two artists, Radiohead and Coldplay, and two cities, Los Angeles
and Seattle. We want to compare how similar Los Angeles' preferences
are to Seattle's. In one sense, they are similar: each city listens to
roughly 50\% more Radiohead than Coldplay. However, if we look at the
absolute number of listens in each city, the cities are far apart
simply because Los Angeles is much larger than Seattle.

In order to compare the similarities of cities regardless of their
size (i.e., last.fm activity level), we always perform Euclidean
normalization on the rows of each listen matrix, which ensures that
each row vector (i.e., each city's listening preference) has the same
length. In other words, the Euclidean normalization puts the row
vectors of each listen matrix on the unit circle, as in
\cref{fig:vec-analysis}(B).  This type of normalization is standard in
the field of Information Retrieval \cite{manning2008}. 

\xhdr{Genres} In the analysis below, it will be important to
distinguish between various genres of music. In order to determine
which genres exist, and which artists belong to each genre, we use
\lfm's \textit{tag} API.  Examples of tags include rock, seen live,
alternative, indie, electronic, and pop (these are the 6 most popular
tags). For each tag, the \lfm{} API also indicates the one thousand
most popular artists that belong to that tag. We construct the listen
matrix associated with a given tag by including only those columns
which represent artists included in the list of top thousand artists
for that tag. We will subsequently refer to the term ``tag'' by the
more conventional term ``genre''. Some tags, e.g. ``seen live'' are
clearly not genres - these are not considered in the analysis
presented here.

\xhdr{Missing data} Inspection of the data indicates that fourteen of
the weeks are outliers in the sense that around the world, little if
any music was listened to. We believe that during these weeks the
last.fm scrobbling service was not operating as usual. In the analysis
below, we omitted from all measurements the contributions that
involved one or more of the missing weeks.

\section{Music knows no borders, yet geographic clusters are strong}
\label{sec:statistics}

\begin{figure}[htp]
  \centering
  \includegraphics[]{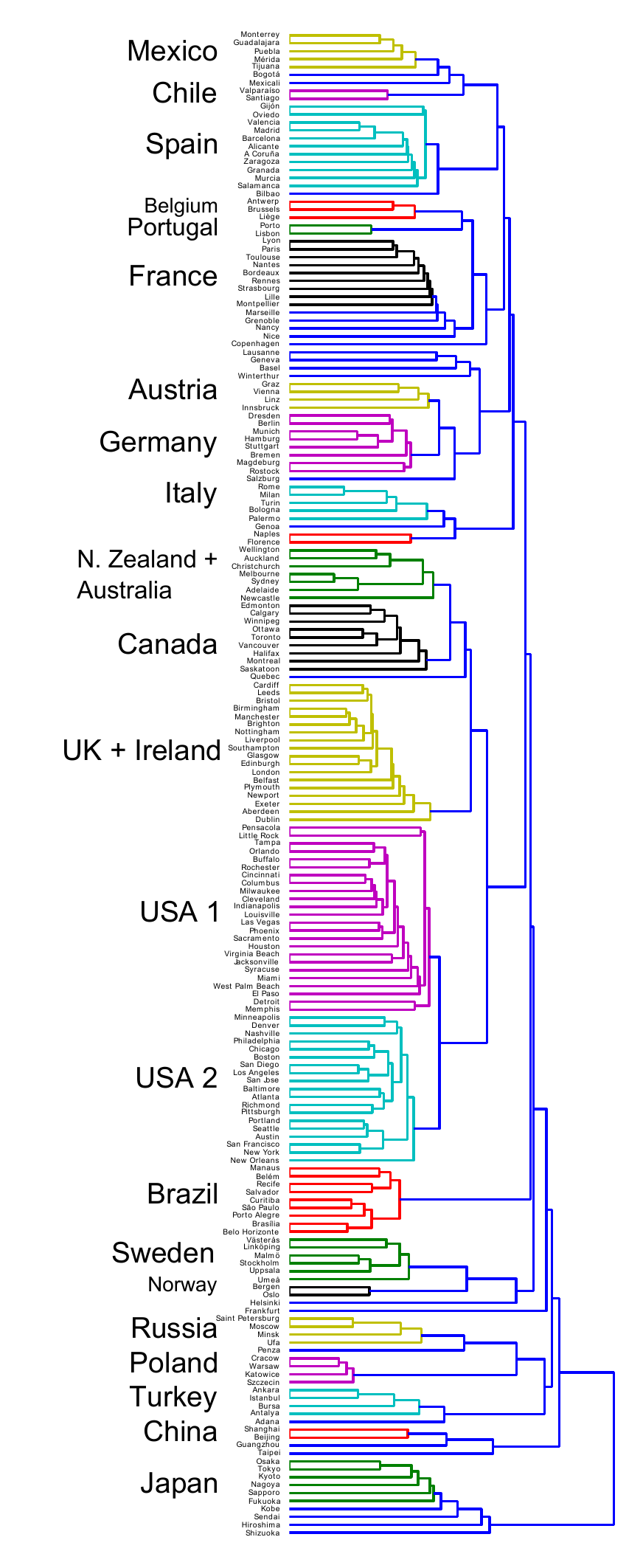}
  \caption{Hierarchical clustering based on average linkage clustering
    of Euclidean distances of cities in the normalized listen
    matrices. (Zoom in electronic version to view city names.)}
  \label{fig:dendrogram}
\end{figure}
Are the listening preferences of \lfm{} users across the world
similar, or do they form coherent clusters? The existence of global
superstars might lead one to believe that largely similar music is
listened to across the world; indeed, in a comprehensive study of the
top-40 music charts of 22 countries, Ferreira and Waldfogel found that
31 artists artists appeared simultaneously on at least 18 countries in
one year \cite{ferreira2010}. Of these 31 artists, 23 were US
American. That such a small set of artists appeared on charts all
around the world suggests a high degree of homogeneity around the
world.

Despite this appearance of global homogeneity, in this section, we
present results which indicate that there are clusters of cities that
have their own idiosyncratic preferences, and that these clusters are
closely related to geographic distance, nationality, and language.

\xhdr{Producing a hierarchical clustering}
To construct the dendrogram shown in \cref{fig:dendrogram}, we
performed average linkage clustering (an agglomerative clustering
algorithm) on a distance matrix $\bf{D}$ of the cities, a square
matrix where each entry $\mathbf{D}_{i,j}$ is the Euclidean distance
between city $i$ and $j$. Instead of constructing the dendrogram based
on just a single listen matrix, we summed together the distance
matrices associated with the all of the listen matrices in our
dataset. The colored clusters are the result of taking a flat
cut to the dendrogram at a height which we chose manually. For an overview of
this type of hierarchical clustering, as well as a description of the
software package we used, see \cite{mullner2011}.

\xhdr{Discussion} If we look at the lowest level structure of the
dendrogram--i.e., the pairs of cities that are most similar to each
other--we observe that every pair involves two cities of the
same nationality. Many of these pairs are composed of cities that are,
in the context of their countries, geographically close to each other:
Cincinnati and Columbus, Portland and Seattle, Berlin and Dresden,
Edmonton and Calgary, Lausanne and Geneva, Gijon and Oviedo,
Birmingham and Manchester, and Edinburgh and Glasgow. However, there
are a few noteworthy exceptions to this trend: New York City and and
San Francisco, Milan and Rome, Munich and Hamburg. These pairs of
cities are close in the dendrogram indicating high relative
similarity. This is surprising because in each case both cities in the
pair have many geographically nearer cities which would seem to be
more likely candidates for most similar counterpart. For example, San
Jose is geographically adjacent to San Francisco, but San Francisco's
users are more similar to NYC's.

At an intermediate level of structure in the dendrogram--the colored
clusters--we see that again, nationality dominates. In the cases
where two countries have been put into the same cluster, such as New
Zealand and Australia or Ireland and the United Kingdom, those two
countries still show separation within their cluster. Intriguingly,
the United States shows two somewhat distinct clusters which are not
geographically coherent and are difficult to explain. We note only
that the USA 2 cluster appears to contain both the largest
metropolitan areas (NYC, Los Angeles, and Chicago), as well as several
cities known for having a large ``hipster'' population, which is
passionate about appearing to know much about music, and likely to use
\lfm{} (such as San Francisco, Austin, Portland, and Seattle).

At the highest level of the hierarchical structure, we observe how the
colored clusters are related to each other. Language seems to be the
key here; Anglophone clusters are closely related, as are the Spanish-speaking
and German-speaking clusters. It is interesting to see that Switzerland, including its
French-speaking cities, are most closely related to German speaking
countries. Chilean cities' musical preferences are more similar
to Mexican cities than to Brazilian cities. We note that although in
this dendrogram Canada appears to be more similar to N. Zealand and
Australia than to the USA, this tendency was not very robust: if we
ran the clustering on a subset of only more active cities, then it was
usually more similar to the US cities. On the other hand, the other
features we have described here were robust in this sense.

\section{Methodology: Detecting leaders and followers}
\label{sec:vector-analysis}
\begin{figure}[htp]
  \centering
  \includegraphics[width=0.98\columnwidth]{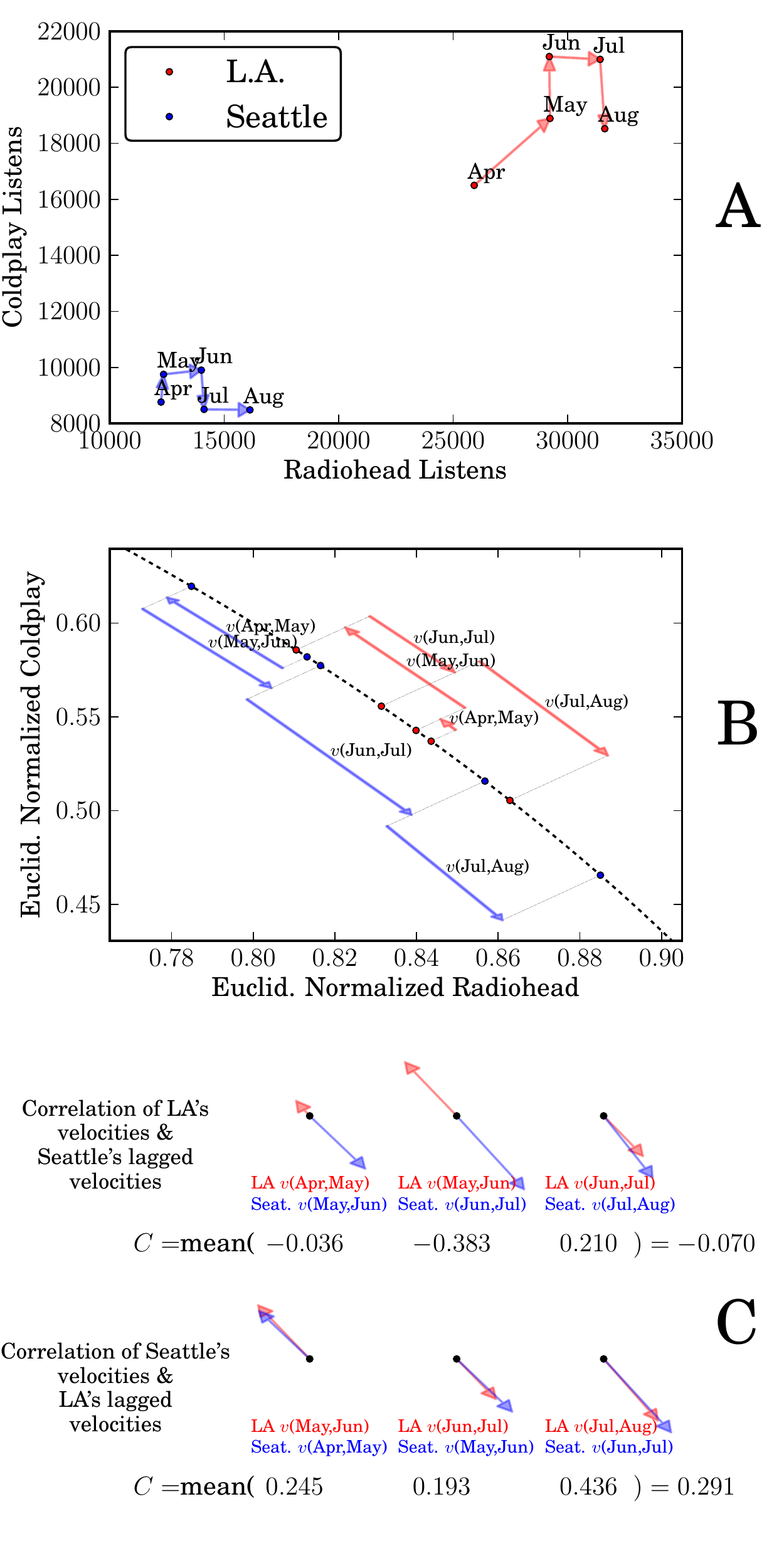}
  \caption{\textbf{Calculating lagged correlations} (here shown with
    imaginary data for ease of explanation). \textbf{(A)} First, for
    each city, we collect from last.fm the number of times that each
    artist was listened to in a given month. \textbf{(B)} To be able
    to compare cities with different levels of last.fm activity, we
    next normalize the number of listens in each city by that city's
    Euclidean norm. We focus on the velocity (change in the normalized
    artist popularity) from the previous month $i-1$ to the current
    month $i$, denoted as $v_j(i-1, i)$ for city $j$, and depicted by
    the arrows in (B). \textbf{(C)} For each pair of cities $(j,k)$,
    we measure the similarity of $v_j(i-1,i)$ and $v_k(i-2,i-1)$ by
    taking the dot product of these velocities.  This yields a list of
    similarities over time; we define the lagged correlation to be the
    mean of these dot products.  In this toy example, it should be
    clear from glancing at the trajectories of Seattle and LA that LA
    is following Seattle, and not the other way; the correlation
    measure presented here successfully indicates this tendency.}
  \label{fig:vec-analysis}
\end{figure}

To detect leader-follower pairs, we adapt the methodology of
Nagy \etal{} \cite{Nagy2010}, which is based on finding lagged correlations, and
was previously applied to finding leadership in pigeon flocks.
In \cref{fig:vec-analysis}, we display some of the key steps of the
method we employ to find leaders and followers. Here we show made-up
data for explanation purposes: we depict a scenario with just two
cities, Los Angeles (LA) and Seattle, and two artists. We are
interested in determining whether
\begin{itemize}
\item LA follows Seattle (in this case we draw the directed edge LA
  $\rightarrow$ Seattle)
\item Seattle follows LA (Seattle $\rightarrow$ LA), or
\item neither leads the other (no edge)
\end{itemize}
If an edge exists, we would also like to assign a weight to that edge
which determines the strength of the leader-follower relationship. We
now explain how we decide on the relationship type and weight.

\xhdr{Calculating lagged correlations}
We begin by performing Euclidean normalization on each city's
listening frequency vector in every listen matrix, as previously
described in \cref{sec:data} and visualized in the change from
\cref{fig:vec-analysis}(A) to \cref{fig:vec-analysis}(B). Each of the
blue arrows in \cref{fig:vec-analysis}(B) is a velocity
$v_{seattle}(t,t+1)$ that represents the change that takes place in
the listening habits of Seattle from one month $t$ to the next month
$t+1$. For example, to find Seattle's velocity from June to July
$v_{seattle}(June, July)$, we subtract Seattle's row in the normalized
listen matrix for time-step June from the corresponding row from the
matrix for July.

As mentioned in \cref{sec:data}, each listen matrix is based on a
four-week window of \lfm{} data, which means that to calculate one of
these velocities, we use eight consecutive weeks (two
four-week windows).  We successively slide this eight week
period one week forward in time, giving us one velocity associated
with each slide. We are left with a sequence of velocities for each
city.

To measure whether Seattle follows LA, we measure the similarity
of each of Seattle's velocities with LA's velocities from one month
earlier, as in the top half of \cref{fig:vec-analysis}(C). We measure the
similarity between two velocities using the dot product (as in
\cite{Nagy2010}). We call the
average of these lagged similarities the \textit{correlation of LA's velocities with
Seattle's lagged velocities}, where the lag size is one month, and we
refer to this measure as $C$.

\begin{figure*}[htpb]
  \centering
  \subfloat[All Music]{\label{america-all}\includegraphics[height=1.1\columnwidth]{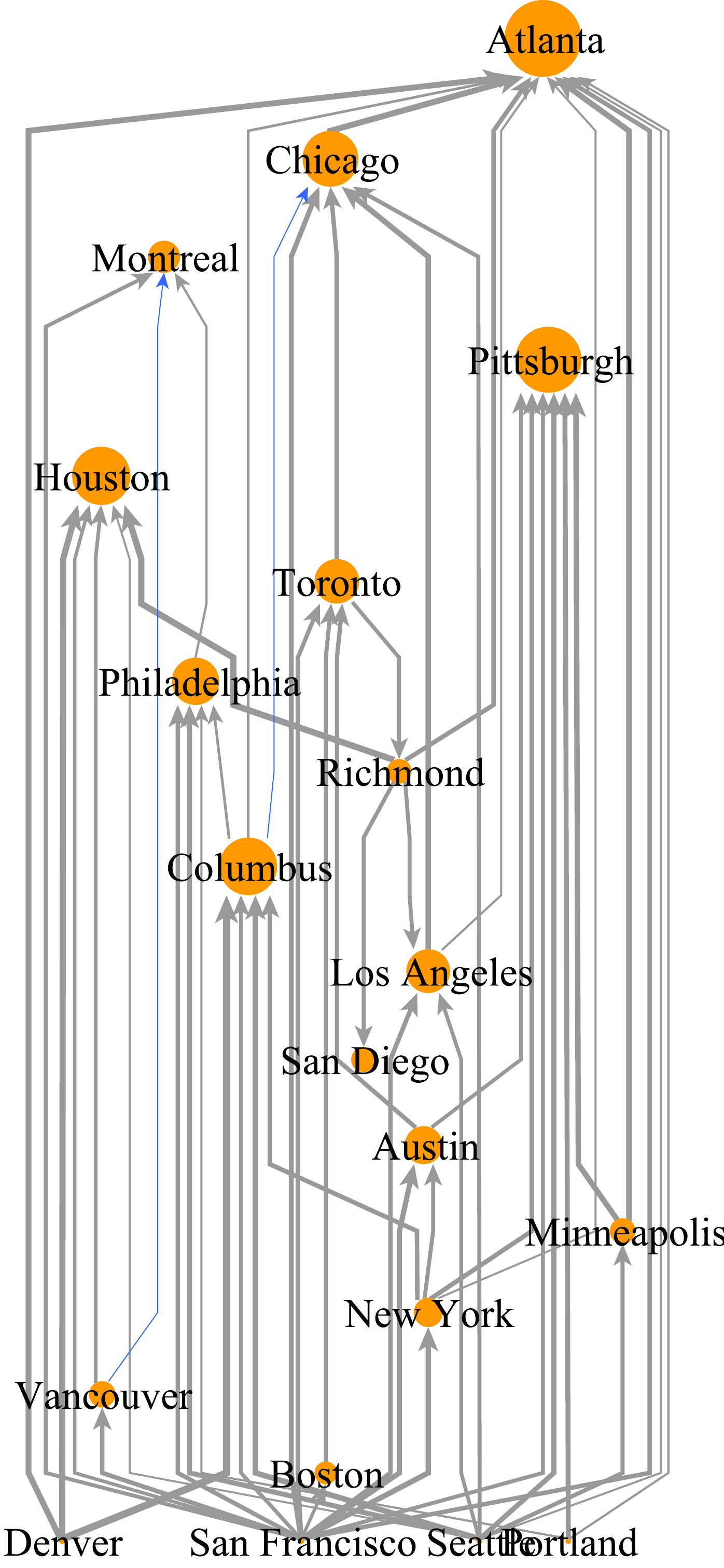}}
  \hspace{0.4cm}
  \subfloat[Indie Music]{\label{america-indie}\includegraphics[height=1.1\columnwidth]{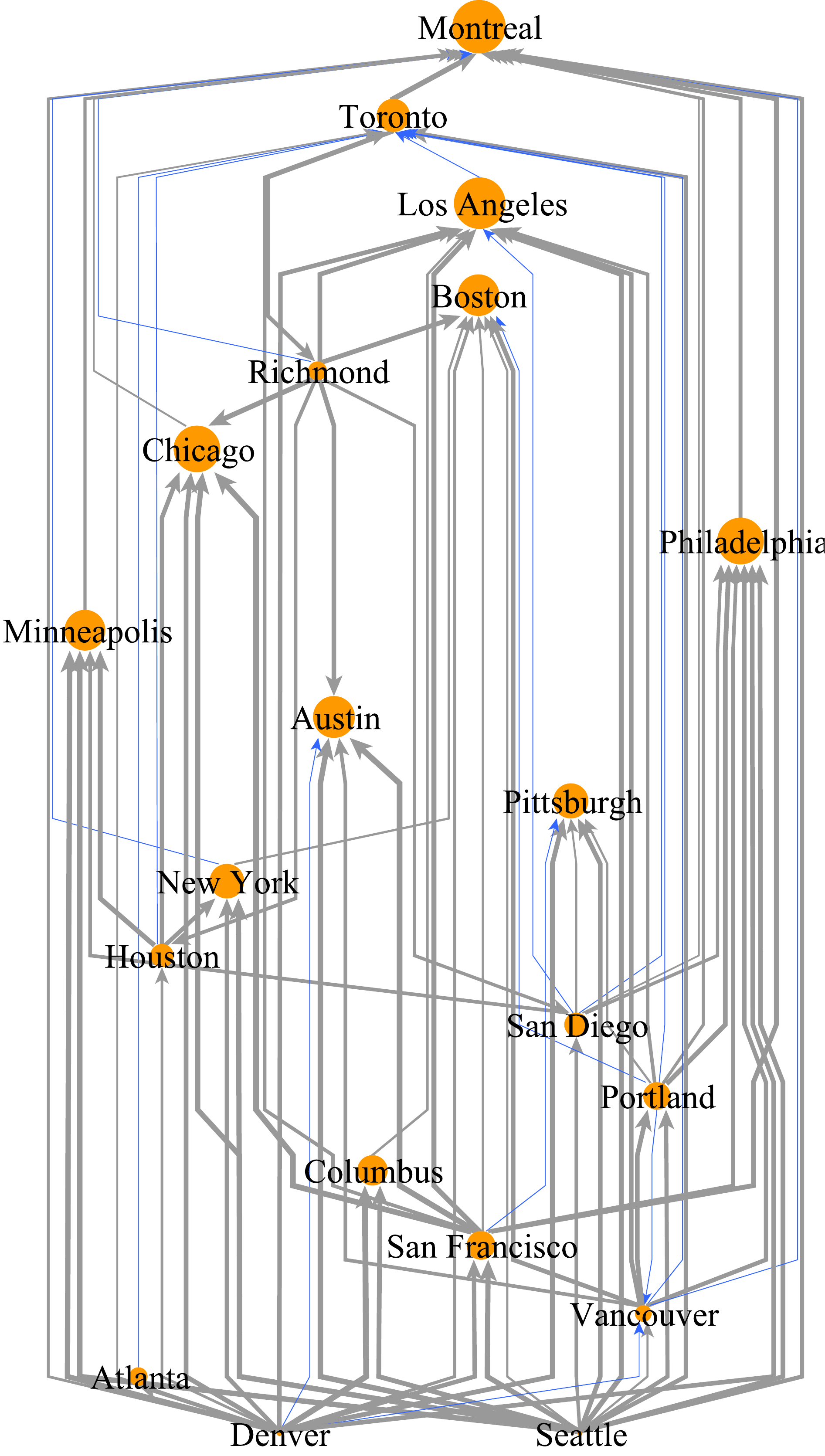}}%
  \hspace{0.4cm}
  \subfloat[Hip Hop]{\label{america-hiphop}\includegraphics[height=1.1\columnwidth]{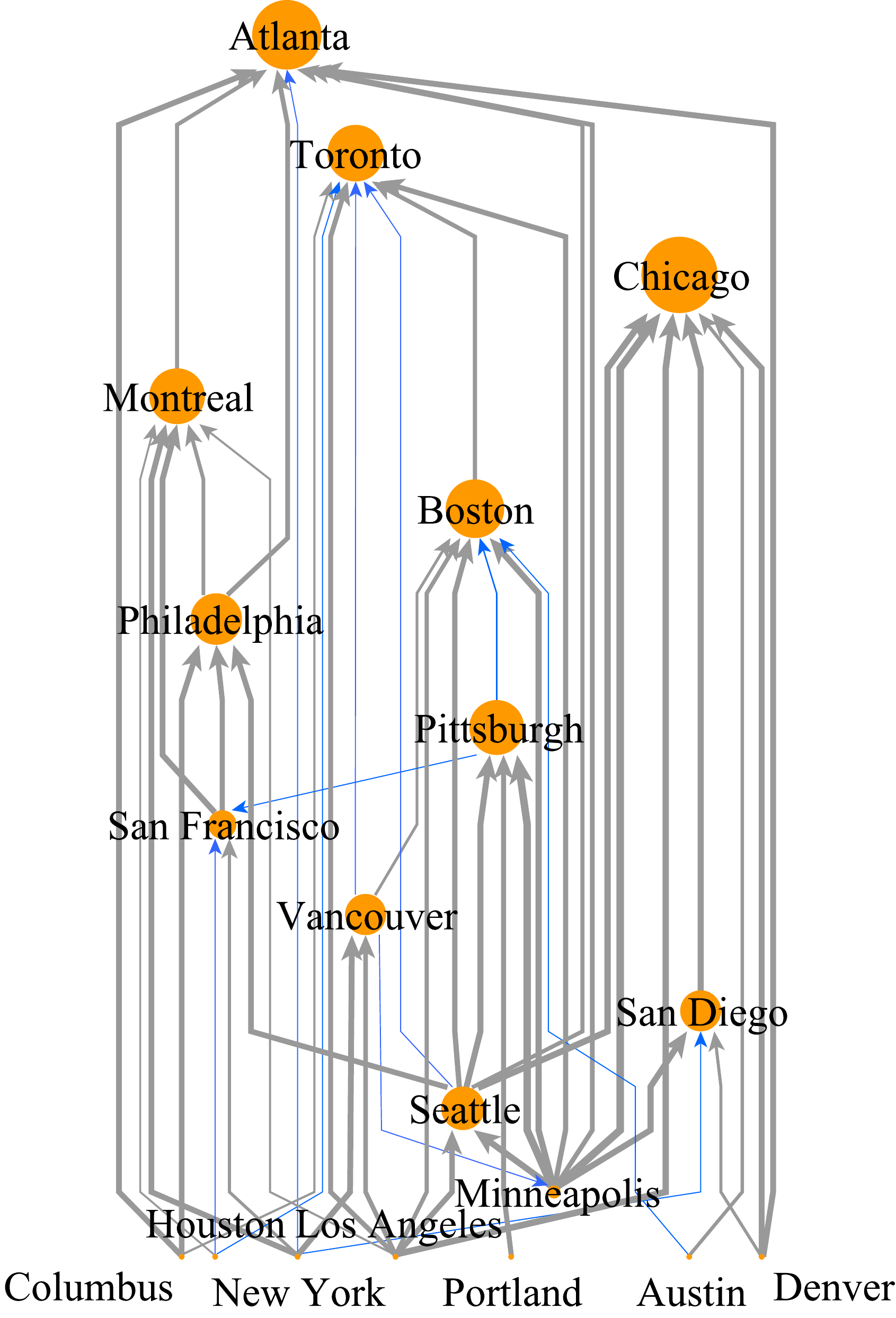}}%
  \caption{Leader-follower network for the twenty most active cities in two
    Canada and the USA. Within each diagram, the height of the nodes
    corresponds to their PageRank, their size to their weighted
    in-degree. Edge width is determined by the correlation, as defined
    in \cref{sec:vector-analysis}. Gray edges have a lag time of 1,
    2, or 3 weeks, blue edges have a lag time of 4 or 5 weeks.}
  \label{fig:flow}
\end{figure*}

In the example displayed in \cref{fig:vec-analysis}, the lag size is
fixed at one month. However, there is no reason to believe that this
lag size should be the same for all dyads and it would be arbitrary to
settle on one month. Along the lines of Nagy \etal, for
each dyad, we consider lag sizes of 1-5 weeks, and we
choose the one which yields the largest correlation. We therefore let
the data decide how this parameter ought to be set. In practice the
lag size which maximizes the correlation tends to be one week, however
there are also cases where the strongest correlation is at four or five weeks (see
blue edges in \cref{fig:flow}) -- in these cases the correlation
tends to be weak.

\xhdr{Deciding which edges to accept} Up to now, the methodology
described in this section closely resembles the one used by Nagy
\etal{} However, we find it necessary to modify their final two steps,
which determine
\begin{enumerate}
\item[\textbf{(1)}] whether a correlation is strong enough to be accepted
\item[\textbf{(2)}] the direction the relationship if one exists.
\end{enumerate}
For step \textbf{(1)}, Nagy \etal{} accept only those leader-follower
relationships which have a correlation above some threshold, either
0.5 or 0.9. This criterion is inappropriate in our case because the
magnitude of the dot products are very small, on the order of 0.01 to
0.001. They are much smaller because, due to the way we normalize
data, cities mostly stand still and move only slightly from week to
week; furthermore we are in a much higher dimensional space. For these
reasons it is hard to pick a threshold for the lowest admissible
correlation size. Instead, we perform one sample t-test on the
distribution of dot products. If we cannot reject the hypothesis that
the the mean of the distribution equals zero, then we say no
leader-follower relationship exists.

It could be the case for a dyad ${i,j}$ that after performing
step \textbf{(1)}, $i$ appears to follow $j$ \textit{and} $j$ appears
to follow $i$. While in this case Nagy \etal{} simply choose the
direction that is larger (even if it is just marginally larger), we
argue that in this situation perhaps neither city is really leading
they other, and instead they are moving together. To make sure there
is a clear direction to the leader-follower relationship, we perform a second
$t$-test to make sure that the two correlations (which are means of
dot products) are not equal; here we
use a two-sided, paired $t$-test. If a one correlation is larger,
then we accept the leader-follower pair associated with that
correlation as a directed edge, otherwise we say no leader-follower
relationship exists.

In the following results, we set $p=0.01$ for all t-tests. We note
that our use of t-tests here is heuristic; for example, we do not test
to make sure that the distribution of dot products is Gaussian
(although they do appear reasonably symmetric and we obtained
qualitatively the same results when outliers were removed), and we
do not correct for our testing of multiple hypotheses. We use the
t-tests as a selection criterion to identify the more pronounced leader-follower relationships, not because we
 rely on their validity in a statistical sense.


\section{Results: The geographic flow of music}
\label{sec:flow}
\begin{figure*}[tpb]
  \centering
  \subfloat[All music]{\label{europe-all}\includegraphics[height=1.1\columnwidth]{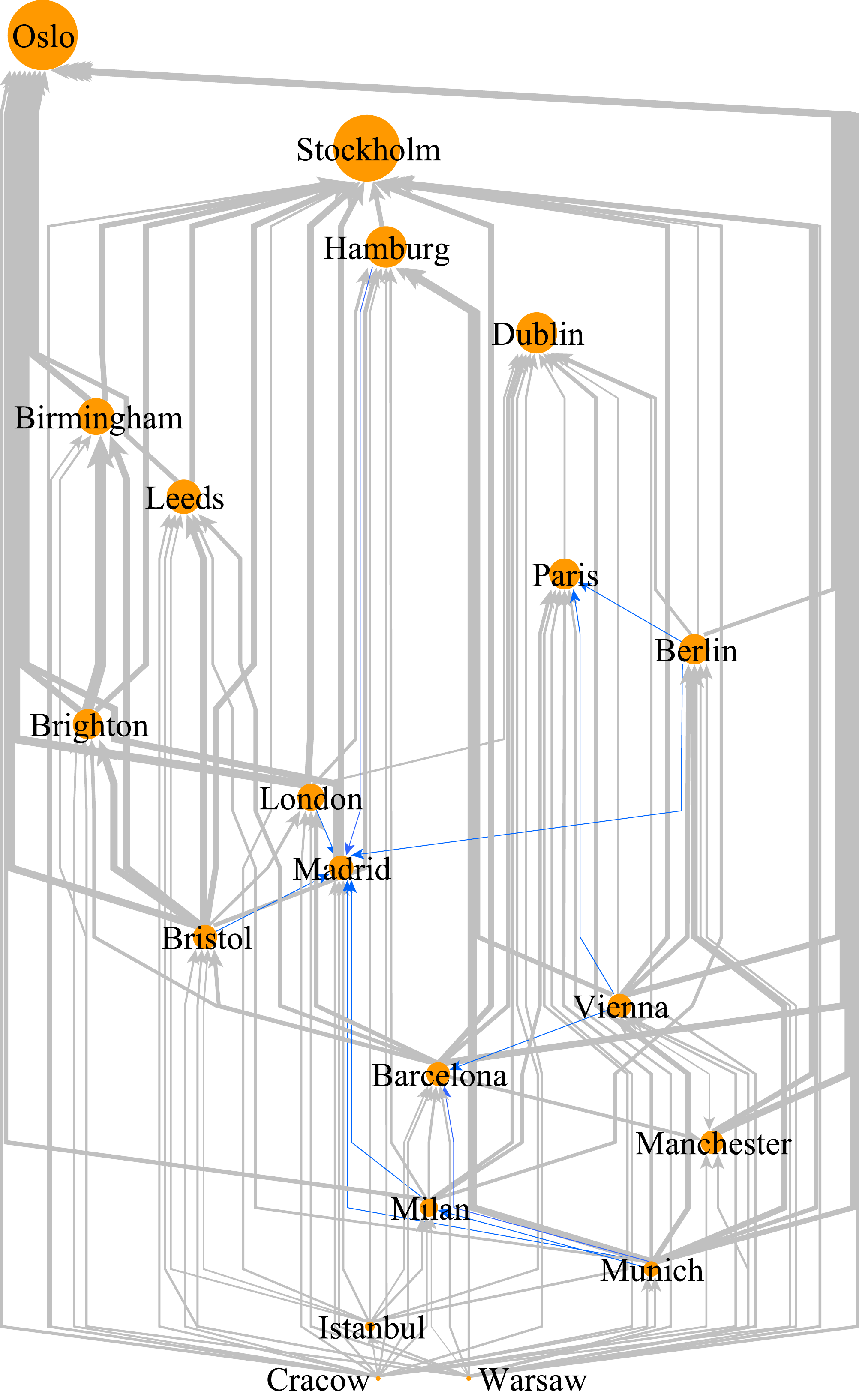}}
  \hspace{1.0cm}
  \subfloat[Indie Music]{\label{europe-indie}\includegraphics[height=1.1\columnwidth]{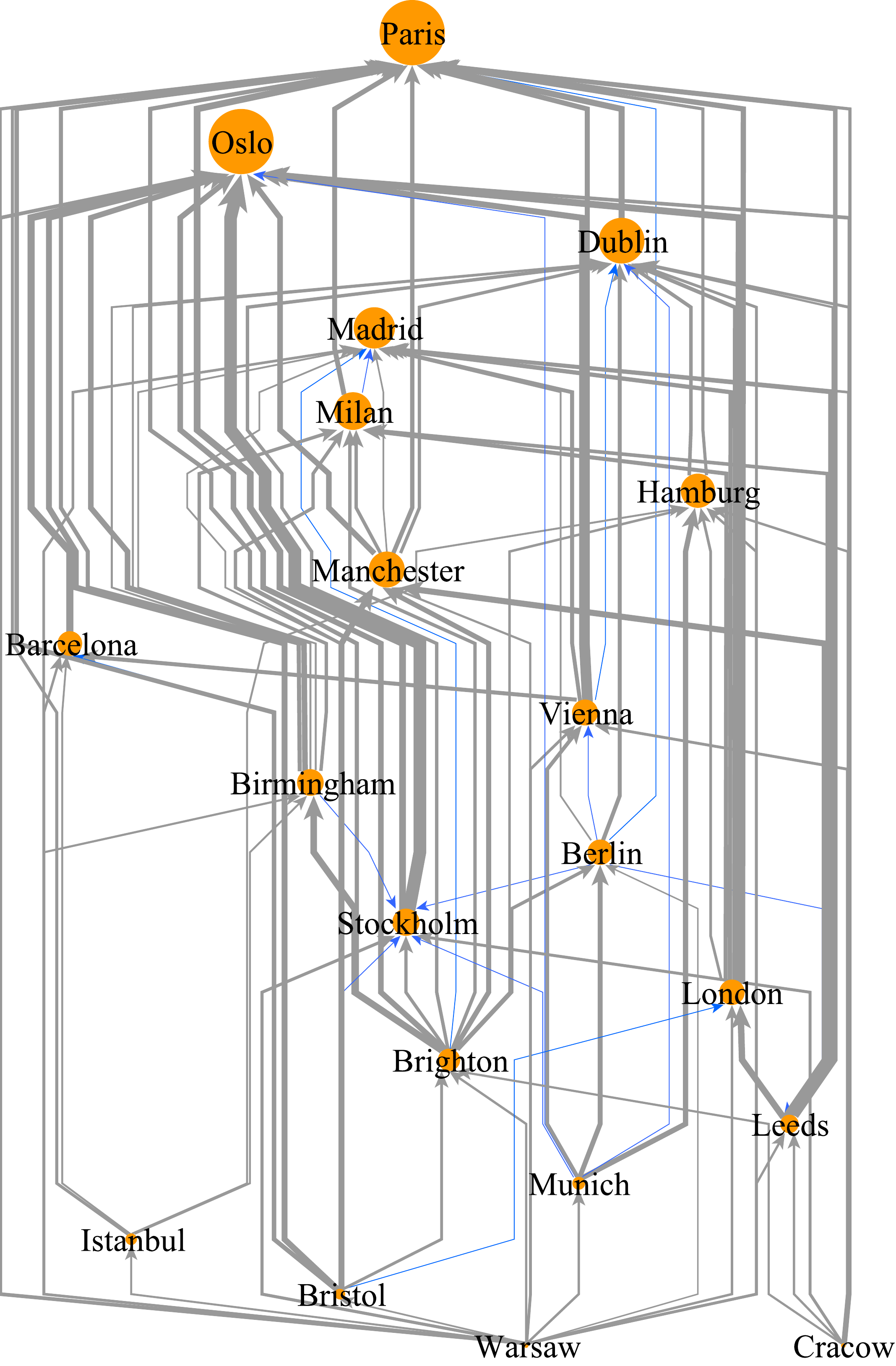}}%
  \caption{Leader-follower network for the most active cities in
    Western Europe. Sizes, positions, and colors as in
    \cref{fig:flow}. Zoom-able online.}
  \label{fig:euro-flow}
\end{figure*}

In the previous section, we described how we determine whether a
leader-follower relationship exists between two nodes. In each study
displayed in \cref{fig:flow}, we take a subset of cities, find all
follower-relationships among them, and plot the resulting network. The
edges point from followers to leaders and are weighted by the lagged
correlation, as defined above.

To create the networks in \cref{fig:flow}, we first choose a genre of
music. While it is possible to create a network showing the flow of
all genres of music, as we have done in \cref{america-all} and
\cref{europe-all}, we find this has a disadvantage: depending upon the
genre that one considers, contradictory relationships may exist. For
example, if we consider hip hop music as in \cref{america-hiphop},
then we see that Atlanta has the most prominent position, whereas if
we consider indie music as in \cref{america-indie}, Atlanta has one of
the least prominent positions. By considering all genres at once as in
\cref{america-all}, these trends get washed out by the
multi-dimensional aspect that genre brings to the data.

\begin{table}[t]
\centering
\caption{Leader-follower networks have few cycles}
\begin{tabular*}{0.8\columnwidth}{llc}
\toprule
Region & Genre & \specialcell{\% Edge weight removed\\ to make acyclic}\\
\midrule
\multirow{4}{*}{N. America}  & All & 0.0\% \\
 & Indie & 1.8\% \\
 & Hip hop & 0.0\% \\
 & Rock & 0.0\% \\
 & Classic Rock & 0.0\% \\
 & & \\
\multirow{4}{*}{Europe}  & All & 0.0\% \\
 & Indie & 0.0\% \\
 & Hip hop & 2.2\% \\
 & Rock & 0.0\% \\
 & Classic Rock & 0.0\% \\
\bottomrule
\end{tabular*}
\label{acyclic-table}
\end{table}

In \cref{fig:flow}, we show the leader-follower relationships between
20 cities in the USA and Canada with the largest number of active 
\lfm{} users. We choose this subset because of the noise associated
with small cities that have insufficient data, and because due to
space constraints it is hard to visualize large networks. The most significant property visible in these graphs
is that they are nearly acyclic; for 
example, \cref{america-all} has no edges, and by removing only three
edges from \cref{america-indie}, the graph becomes acyclic. In
\cref{acyclic-table}, we show that this property holds true for the
leader-follower networks created from diverse geographic regions and
genres. To calculate the measure displayed in that table, we first
computed the feedback arc (edge) set, which is the smallest set of
edges that, when removed from a graph, make the graph acyclic. We
measured the percent of the graph's total weight in the feedback arc set.

We believe that this lack of cycles is not an
artifact of our methodology, which focuses only on dyads and does not
consider the network as a whole. Rather, we believe that the lack of
cycles is inherent in the data itself, indicating a clear direction in
the flow of music preferences. Others have argued that a system with a strong leadership
hierarchy ought to be nearly acyclic\cite{Nagy2010,mones2012} so the lack of cycles in our networks is a clear validation of the methodology. 

There are many centrality measures that could be used as criteria for
deciding which cities are the most cutting edge and which are
laggards.  The networks in \cref{fig:flow,fig:euro-flow} display two of
these centrality measures: their height reflects their PageRank, which
seems appropriate  because PageRank is designed
to rank importance of nodes on weighted, directed networks on which a
dynamic process takes place \cite{page1999}. The area taken up by each node reflects
its weighted in-degree. While it is apparent that the PageRank and
weighted in-degree are highly correlated, in some cases they order
nodes differently---for example, in \cref{america-hiphop}, Atlanta has
the largest PageRank, but Chicago has the largest weighted
in-degree. These visualizations were created using the ``status''
layout algorithm of the network visualization software
Visone \cite{brandes2004}.

For us, the most surprising features of \cref{fig:flow} are (1) the
middle ranking positions of some of the largest cities, such as NYC and LA
in \cref{america-all} and NYC and Chicago in \cref{america-indie} and
(2) the prominent position of Canadian cities, especially in
\cref{america-indie}. While Montreal is known for having produced some popular
indie bands (such as Arcade Fire and Wolf Parade), this does not
necessarily mean that \lfm{} listeners from Montreal would be
generally leaders in their taste in indie music; in any case, New York
City is presumably home to more prominent indie artists than Montreal.

While the diagrams in \cref{fig:flow} display the leader-follower
relations for a relatively homogeneous cultural region, those in
\cref{fig:euro-flow} display these relations in Europe, a region more
culturally and linguistically diverse. It is interesting to note that
many of the most heavily weighted edges are between cities in
different countries and which speak different languages. For example,
London, Birmingham, Brighton, and Bristol, have a much stronger follower
relationship with Oslo and Stockholm than with each other (London's unremarkable
position is also noteworthy). Similarly, Cracow and Warsaw do not
follow each other, rather their strongest edges point to German and
Scandinavian cities.

Along the lines of this last observation, it is noteworthy that in
general many of the edges with the largest weights 
connect cities which were not similar to each other in the
hierarchical clustering in \cref{fig:dendrogram}. For example, the
Canadian cities are located far away from the US cities in that
clustering, yet here there is a strong flow from
the former to the latter. Although pairs of cities such as Portland and
or NYC and San Francisco are very similar in the clustering, they are
connected in \cref{america-indie} by only weak edges. One speculative
explanation is that cities which have very
similar listening habits are largely synchronized with each other, and
therefore there is little potential for novel information to flow between them. 
For example, the leading city in \cref{america-indie}, Montreal, is
unique in that the language spoken by the majority is not English but
French, a difference which may provide it with novel information.


\section{Hypothesis: large cities are leaders}

\label{sec:superlinearity}

As noted in the introduction, there is currently much excitement
surrounding the observation that productivity, efficiency, and
innovation all scale super-linearly with the size of a city.  For an
accessible, high-level overview of this discussion, see
\cite{bettencourt2011}; for extensive empirical evidence for the
universality of this relationship, see \cite{bettencourt2007}; and for
a proposed causal mechanism, see \cite{arbesman2011}.

This work makes many fascinating empirical observations as well as an
interesting comparison between organisms and cities; here we summarize
only a few main points. The first is that the total productivity
of a city $P$ is super-linear. In data collected so far, a power-law
relationship appears to provide a reasonable fit, so that total production
in an $N$ person city is well approximated by the relationship
$P(N)=N^{\beta}$, where $\beta \approx
1.2$. \cite{bettencourt2007} This means, for example, that a person
living in cities with 10 million inhabitants is roughly 2.5 times as
productive (in terms of wealth production, creativity, patents, and
other measures) as an individual living in a city with only 100 thousand
inhabitants. Consumption of water, gasoline, or electricity appear to
have a linear relationship, so people in smaller and larger cities
consume the same amounts. Certain types of infrastructural needs, such
as the number of gasoline stations, the meters of electric cabling
installed, and road surface area, increase sub-linearly, with the
scaling exponent $\beta\approx0.8$, indicating economies of scale.

Bettencourt \etal{}, the authors of \cite{bettencourt2007}, also
suggest that the very pace of life in large cities is faster, and Arbesman et al. 
\cite{arbesman2011} propose that productivity gains may be attributed
to the increased probability of ties between diverse groups, which
helps information spread quickly. If the pace of life in larger cities
were faster, and the spread of information more efficient, then it
would be reasonable to expect that larger cities would lead smaller
ones in adopting fresh music and abandoning stale music. Here we test
this hypothesis by measuring whether city size is positively and
strongly correlated with a position of leadership in the network flow
diagrams presented in \cref{sec:flow}.

Bettencourt \etal{} are careful to treat each ``national urban system''
separately, because otherwise their measurements might be confounded
by the fact that different countries have economies at different
levels of development. Thus,  they do not expect that all cities
around the globe which are of the same size should have the same level of
production; rather, they  expect this only within a tightly
integrated economic region. (They do however argue
that the same scaling exponent exists in every nation.) The North
American cities in \cref{fig:flow} belong to a tightly integrated
economic area at a similar level of development, so we test this
hypothesis on that set of cities. For US population sizes, we use the US Metropolitan
Statistical Areas, as Bettencourt \etal{} (although we use the newer
data from 2010), for Canadian population sizes we use the Census
Metropolitan Areas from 2011.

In \cref{superlinear-table} we display some measures of the
relationship between the population size of a city and its leadership
status in the diagrams depicted in \cref{fig:flow}. The second and
third columns display the Spearman's rank correlation of population
size and PageRank, and population size and the weighted in-degree,
respectively. The final column
shows, for all edges, the percentage of the total weight that comes
from edges where the larger city is the leader.

\begin{table}[t]
\centering
\caption{Relationship between a city's population size and status in network}
\begin{tabular*}{0.8\columnwidth}{lccc}
\toprule
&\multicolumn{2}{c}{\specialcell{Spearman rank correlation\\of centrality \& population}}\\
\cmidrule(lr){2-3} 
Genre & PageRank & In-degree & \specialcell{\% Edge weight where\\leader larger}\\
\midrule
All  & 0.34 & 0.18 & 55\% \\
Indie & 0.61 & 0.61 & 61\% \\
Rock  & 0.21 & 0.26 & 63\% \\
Hip Hop & 0.38 & 0.28 & 59\% \\
\bottomrule
\end{tabular*}
\label{superlinear-table}
\end{table}

While these correlations between city size and leadership position are positive, most of these
relationships are quite weak when compared with those observed in the
above-mentioned work on superlinearity of cities.
We were surprised that they were not stronger. In most
genres, the Spearman correlation coefficients are smaller than we
expected, and the percentage
of edge weight that comes from edges where small cities are led by
larger cities is not very far from 50\%. Additionally, although there
are some cities in North America which dwarf most of the other large
cities (such as NYC, with 18.9 million residents, and LA with 12.8
million), in many cases these cities do not occupy prominent positions
in \cref{fig:flow}.

Indie music is an exception--here the correlation between leadership
and city size is quite large. We are not sure why this is the
case--perhaps this genre is quicker moving or more urban than the
others (although presumably hip hop is also quite an urban genre).

The work on scaling laws in cities which we have summarized in this
section is significant because it appears to have uncovered a
universal law in the social sciences, one which can make quantitative
predictions. Our point here is not to claim that our results
contradict this law. The preliminary results presented in this paper
suggest that, in the specific context of being cutting edge in music,
cities are idiosyncratic. Larger cities are not predictably and
generally ahead of smaller cities. In other words, a city is more than
the number of its inhabitants, it might lead the trends in one genre
while lagging them in another.

\section{Discussion and Future work}
\label{sec:conclusion}
One major question hangs over the results presented above: why should
we believe that our models of flow, as pictured in the network
diagrams displayed in this paper, are valid? On the one hand, two
aspects of our methodology lend the results credibility: that each
leader-follower relationship underwent a $t$-test, and that when all
of the leader-follower relationships were put together into a graph,
they formed directed acyclic graphs, which indicate a direction of
flow in a strict sense. For these two reasons, our method
distinguishes itself from other unsupervised methods---such as many
clustering methods---which are problematic because they return results
regardless of whether there is structure in the underlying data. In
other words, if we shuffle our data around so that random noise
dominates any signal of leader-follower relationships, our method no
longer detects leader-follower relationships.

On the other hand, certain doubts remain, and we should stress that
our results reflect a work in progress.  For example, a relationship
can be statistically significant but at the same time have a very
small magnitude. We would be more confident of our results if we could
demonstrate that the model that we create is meaningfully
predictive. That is, given our model of leader-follower relationships
among cities, and given a record of past listening behavior, we should
be able to predict the changes in listening behavior that will occur
in the near-term future better than a reasonable baseline
predictor. We have not yet demonstrated that our models have this
predictive power, although we plan to attempt this validation in
future work. 


\section*{Acknowledgment}
This material is based upon works supported by the Science Foundation
Ireland under Grant No. 08/SRC/I1407: Clique: Graph \& Network
Analysis Cluster.


\bibliographystyle{IEEEtran} \bibliography{IEEEabrv,refs.bib}

\end{document}